\documentclass{aa}
\usepackage{graphics,latexsym,psfig}
\def\degr{\hbox{$^\circ$\ }}

\def\arcsec{\hbox{$^{\prime\prime}$\ }}

\def\farcs{\hbox{$.\!\!^{\prime\prime}$}}

\def\Ha{\hbox{H$\alpha$\ }}
\def\He2{\hbox{He {\sc ii} $\lambda$4686}}

\def\RL1{\hbox{{R$_{\rm L_{\rm 1}}$}}}
\def\kms{\hbox{km s$^{-1}$}}
	
\begin{document} 
\renewcommand{\textfraction}{0.0}
\title{A spectrophotometric study of RW Trianguli} 
\author{P.J. Groot \inst{1,2}
\and R.G.M. Rutten\inst{3}
\and J. van Paradijs \inst{2,4}}	
\institute{Department of Astrophysics, University of Nijmegen,
  P.O. Box 9010, 6500 GL, Nijmegen, The Netherlands
\and 
Astronomical Institute `Anton Pannekoek'/ CHEAF, Kruislaan 403, 
1098 SJ, Amsterdam, The Netherlands
\and Isaac Newton Group of Telescopes, Apartado de Correos 321,
E-38700, Sta Cruz de La Palma, Islas Canarias, Spain 
\and Physics Department, UAH, Huntsville, AL 35899, USA}
\date{received date; accepted date} 
\offprints{pgroot@astro.kun.nl}

\abstract{ On the basis of spectrophotometric observations we
reconstruct the accretion disk of the eclipsing novalike cataclysmic
variable RW Tri in the wavelength region 3600-7000\AA. We find a
radial temperature profile that is, on average, consistent with that
expected on the basis of the theory of optically thick, steady state
accretion disks and infer a mass-accretion rate in RW Tri of
$\sim10^{-8}$ M$_{\odot}$ yr$^{-1}$. The line emission is dominated by
two areas: one around the hot-spot region and one near the white
dwarf. Both emission regions have appreciable vertical extension, and
seem to be decoupled from the velocity field in the disk. In our
observations RW Tri shows a number of features that are characteristic
of the SW Sex sub-class of novalike stars. The appearance of a
novalike system as a UX UMa/RW Tri or SW Sex star seems to be mainly
governed by the mass-transfer rate from the secondary at the time of
observation. 
}

\maketitle

\section{Introduction \label{sec:introrwtri}}
RW Tri is one of the earliest known cataclysmic variables (CVs),
discovered by Protitch (1937). It has been studied extensively
photometrically, but has been largely neglected in spectroscopic
studies. To our knowledge only two extensive optical
spectrophotometric studies, by Kaitchuck et al. (1983) and Still et
al. (1995) have been made of this system. RW Tri is generally assumed
to be a standard novalike CV (see Warner, 1995 for a general overview
of CVs), but both spectroscopic studies have shown that the phase
depedence and the light curves of the emission lines show features
that are difficult to explain in a standard CV picture.

RW Tri was included in the broad-band photometry eclipse mapping
studies of Rutten et al. (1992) who showed that the radial temperature
profile of its accretion disk is consistent with the $T\propto
R^{-3/4}$ dependence expected on the basis of accretion disk theory
(see e.g. Frank et al. 1992). RW Tri is in this respect
similar to UX UMa that has been shown spectrophotometrically to follow
the same temperature profile (Rutten et al., 1993; 1994).

RW Tri is also known to undergo irregular brightness variations of up
to one magnitude in its out-of-eclipse brightness, as was first shown
by Walker (1963).  It has even been observed (Still et al., 1995) to
be more than three magnitudes brighter than its `quiescent' value at
AB$\sim$13.5. This irregular behaviour is not unique for RW Tri (see
e.g. the recent results on GS Pav; Groot et al., 1998), although it is
the best documented case.

\section{Observations \label{sec:obsrwtri}}
On the nights of 22-26 October 1994, we obtained a total of 671
low-resolution (8\AA) spectra using the Intermediate Dispersion
Spectrograph with the R300V grating and a 1k$\times$1k TEK CCD,
attached to the 2.5m Isaac Newton Telescope on the island of La
Palma. A wide slit (2\farcs5) and a second star on the slit
(48\arcsec\ NW of RW Tri) were used to obtain differential
photometry. For absolute flux calibration the spectral flux standard
BD +28 4211 (Oke, 1990) was used with a 5\arcsec wide slit for the
standard as well as for RW Tri and its comparison star.

All spectra were obtained with a 50s on-target integration time. With
a $\sim$60s dead-time for CCD readout and data storage, we obtained an
effective time resolution of 110s, or 1/182nd of the orbital period. 
A total of five eclipses were observed. Throughout the nights
CuAr arc spectra were taken for wavelength calibration. All data were
reduced in the standard fashion using the ESO-MIDAS package, with
additionally written software and they were optimally extracted
(Horne, 1986). 

Based on the colour excess given by Rutten et al. (1992) of
$E(B-V)$=0.1 we have dereddened all our spectra using the galactic
reddening coeffcients given by Cardelli et al. (1990),
assuming a standard $R_V$=3.1. 

\section{Ephemeris and System Parameters}

We have phase folded all spectra using the ephemeris given by
Robinson et al. (1991). Trial eclipse maps using the
system parameters given in Table\ \ref{tab:sysrwtri} showed a phase
shift in the phase of minimum light. Shifting the phases by \mbox{--0.0046} of
an orbital period, as has been found before by Smak (1995), corrected for
this. A revised ephemeris is given in Eq.\ \ref{eq:ephrwtri}

\begin{equation}
HJD_{\rm mid\_ecl} = 2\,441\,129.36380(10) + 0.231883297 E
\label{eq:ephrwtri}
\end{equation}

The system parameters of RW Tri are rather uncertain, especially the
($q,i$) pair, where $q$ is the mass ratio ($M_2$/$M_1$) and $i$ the
orbital inclination. Values for $i$ range between 67\degr (Kaitchuck
et al., 1983) to as high as 80\degr (Mason et al., 1997),
with the component masses varying accordingly. Previous studies agree
that the mass ratio is close to unity. We have chosen to use the
values as given in Rutten et al. (1992, see also Table\
\ref{tab:sysrwtri}).

\begin{table}
\caption[]{System parameters of RW Tri.
\label{tab:sysrwtri} }
\begin{tabular}{ll}
Period  & 20034.717 s\\
M$_{\rm WD}$ & 0.7 M$_{\odot}$\\
M$_{\rm sec}$ & 0.6 M$_{\odot}$\\
Inclination & 75\degr\\
Distance     & 330 pc\\
\end{tabular}
\end{table}     

\begin{figure}[htb]
\centerline{\psfig{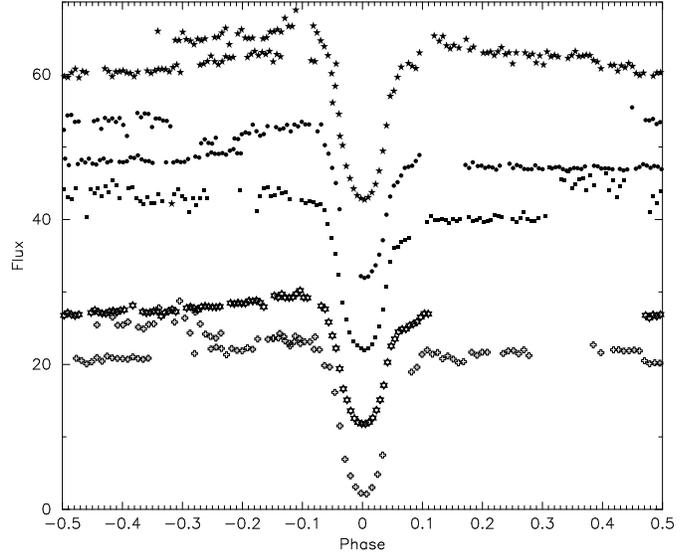}}
\caption[]{The continuum light curves of RW Tri between 4400\AA\ and
4600\AA, from bottom (22 October) to top (26 October). The light
curves are offset by steps of 10 mJy per night, with no offset for the
bottom curve. \label{fig:contlightrw}}
\end{figure}

\begin{figure*}[htb]
\centerline{\psfig{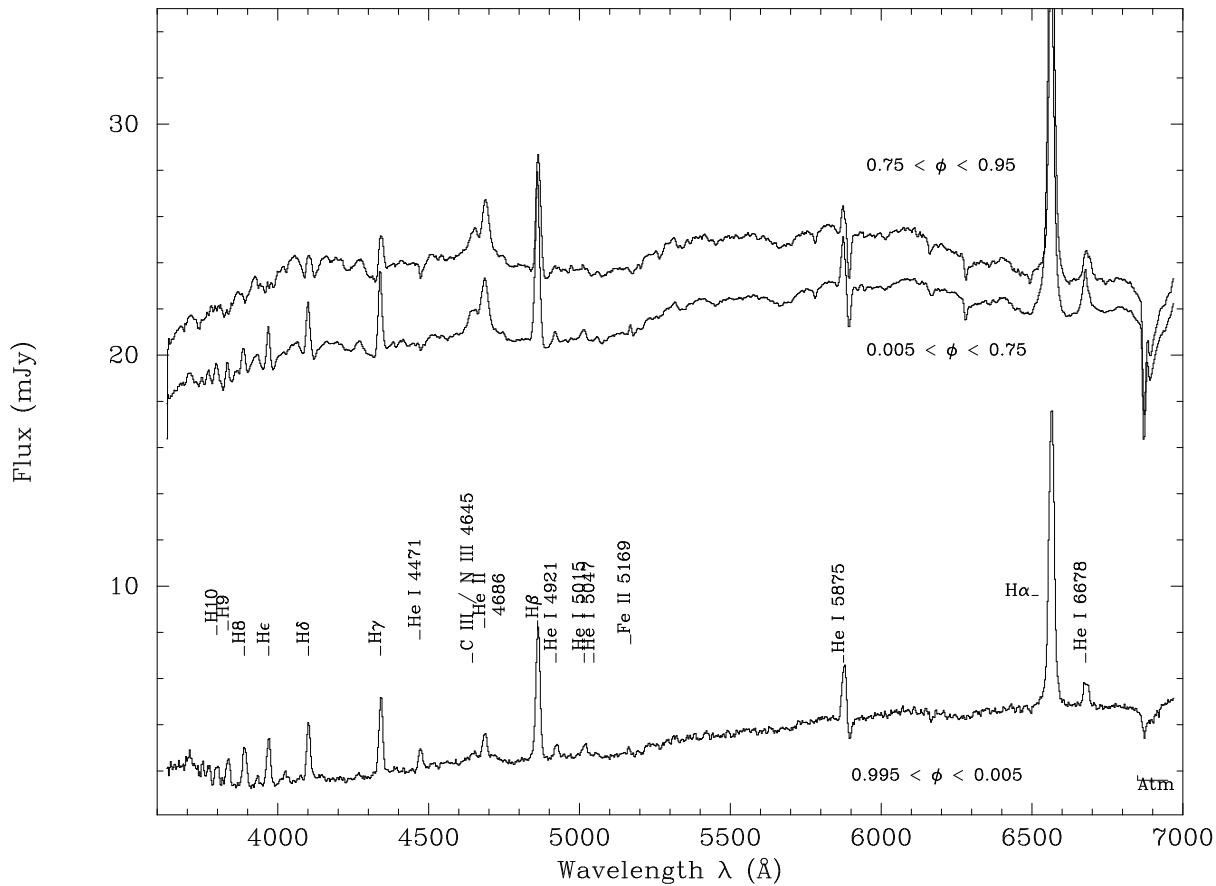}}
\caption[]{The average spectrum of RW Tri, divided in three phase
intervals. The bottom curve shows the spectrum in mid-eclipse
(0.995$<\varphi<$0.005), the middle curve the spectrum outside eclipse
and outside the phase interval that a hot-spot can be visible
(0.005$<\varphi<$0.75) and the top curve shows the spectrum during the
hot-spot phase (0.75$<\varphi<$0.95). 
\label{fig:averwtri}}
\end{figure*}

\begin{figure*}[htb]
\centerline{\psfig{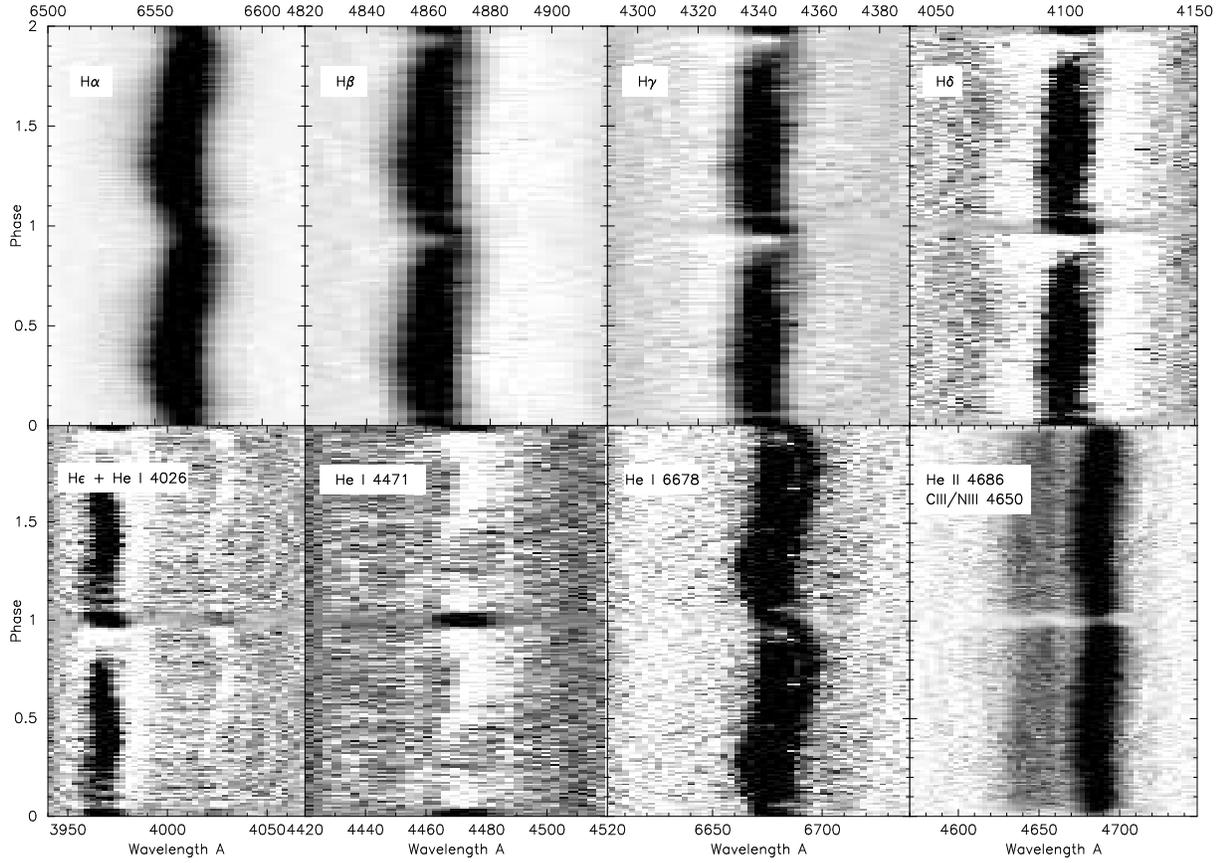}}
\caption[]{The trailed spectra (in bins 0.01 in phase) for \Ha,
H$\beta$ (plus He {\sc i} $\lambda$4921, 
H$\gamma$ and H$\delta$ on the top row and H$\epsilon$ (plus
He {\sc i} $\lambda$4026), He {\sc i} $\lambda$4471, He {\sc i}
$\lambda$6678,  and \He2 plus C{\sc iii}/N{\sc iii} on the bottom row. 
Black indicates emission.
\label{fig:trailrwtri}}
\end{figure*}

\section{Continuum light curves \label{sec:contrwtri}}

The continuum light curves of RW Tri (Fig.\ \ref{fig:contlightrw})
show that the system varied up to $\sim$30\% in its out-of-eclipse
light level from night to night. These short term variations of RW Tri
have been long known (Walker 1963, Smak 1995).  In our observations
the system varied, at 4500\AA, between AB=13.2 (18.5 mJy) on the night
of October 23, to AB=12.9 (23 mJy) on October 26.

The light curves show the eclipse as the most prominent feature. As in
other novalike systems, only a weak orbital hump is seen before the
eclipse. The eclipse shape itself, however, does show a prominent
hot-spot egress feature.

\section{Spectral line behaviour \label{sec:lines}}

\subsection{Average Spectrum \label{sec:avespecrwtri}}

In Fig.\ \ref{fig:averwtri} we show the average flux calibrated
spectrum of RW Tri during our observations. The spectrum shows the usual
emission lines of H {\sc i}, He {\sc i} and He {\sc ii}.
All lines appear single
peaked in our low-resolution spectra. 
In the average, non hot-spot, non mid-eclipse, spectrum (middle curve
in Fig.\ \ref{fig:averwtri}) we can see absorption troughs underlying
the Balmer series, especially well visible in the higher members,
H$\gamma$ and higher. These troughs become deeper when the hot-spot is
visible (top spectrum in Fig.\ \ref{fig:averwtri}).  The red part of
the spectrum shows the Earth's atmospheric features at 6300\AA\ and
6900\AA.

\subsection{Trailed spectra \label{sec:trailrwtri}}

In Fig.\ \ref{fig:trailrwtri} we show the trailed spectra of the
main lines in RW Tri after having subtracted the continuum
contribution by making a linear fit to the adjacent continuum.

In Fig.\ \ref{fig:trailrwtri} we see that most of the lines consists
of at least two components, best seen in the He {\sc i} $\lambda$6678
line. A component with very low radial velocity amplitude dominates
the lower Balmer (H$\alpha$, H$\beta$), the \He2 and lower He {\sc i}
lines. In all these lines a second component, with higher radial
velocity is also seen. It is this second component that is the {\sl
only} one seen, and in absorption, in the He {\sc i} $\lambda$4026 and
$\lambda$4471 lines. The lower He {\sc i} lines show this component in
emission. We will use the \Ha and \He2 lines to derive the phasing and
amplitude of the low-velocity component since this component dominates
in these lines. The He {\sc i} $\lambda$4471
absorption line will be used to derive the radial velocity and phasing
of the higher velocity component. We will here make the assumption
that the amplitude and phase of these two components is the same in
all the lines.

\subsubsection{Radial velocity curves}

\begin{figure}[htb]
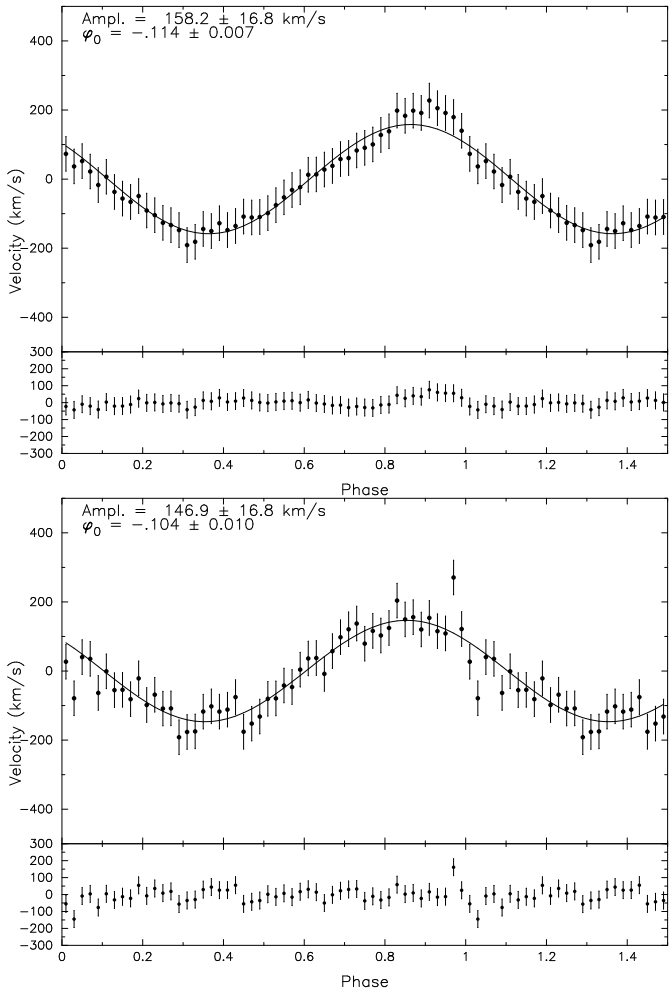

\centerline{\psfig{figure=h1792f5a.ps,angle=-90,width=8.8cm}}
\centerline{\psfig{figure=h1792f5b.ps,angle=-90,width=8.8cm}}
\caption[]{The radial velocity curve of \Ha (top) and \He2 (bottom)
and a sinusoidal fit to the phase interval 0.1$<\varphi<$0.9.
\label{fig:radvelHaHe2}}
\end{figure}

We have used an auto-correlation technique to determine the radial
velocity curve of the low-velocity component dominating \Ha and \He2.
In Fig.\ \ref{fig:radvelHaHe2} we show the radial velocity curve and a
sinusoidal fit to the phase interval 0.1$<\varphi<$0.9. Both \Ha and
\He2 show the same phase lag and radial velocity amplitude, indicating
that they are formed in the same region. The phase lag shows that the
emission site is not on the line of centers, connecting the secondary,
the center-of-mass of the system and the white dwarf.

\begin{figure}[htb]
\centerline{\psfig{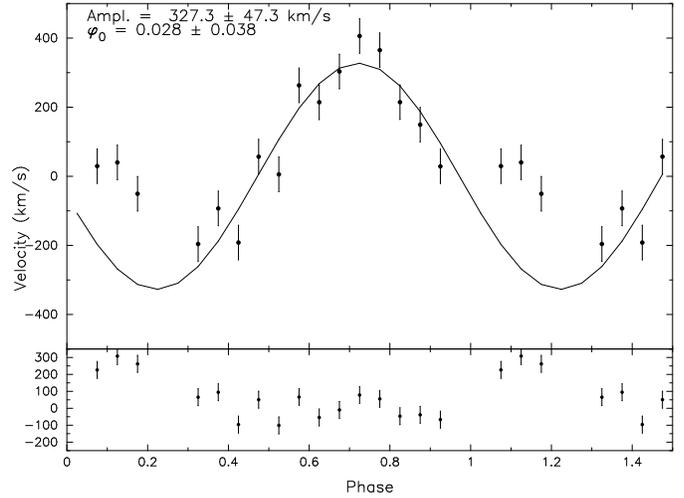}}
\caption[]{The radial velocity curve of He {\sc i} $\lambda$4471,
determined by autocorrelation with the profile at
$0.65<\varphi<$0.7. The sinusoidal fit has been made to the phase
interval 0.35$<\varphi<$0.9. \label{fig:radvelHeI}}
\end{figure}

The radial velocity curve of He {\sc i} $\lambda$4471 (Fig.\
\ref{fig:radvelHeI}) shows that the high-velocity component is in
anti-phase with the secondary star and has an amplitude of $\sim$330
\kms. 

Smith et al. (1993) have shown the secondary
to be a K7-type star. Among the most prominent lines in late K-type
stars is the Ca {\sc i} triplet at 6200\AA. 

\begin{figure}[htb]
\centerline{\psfig{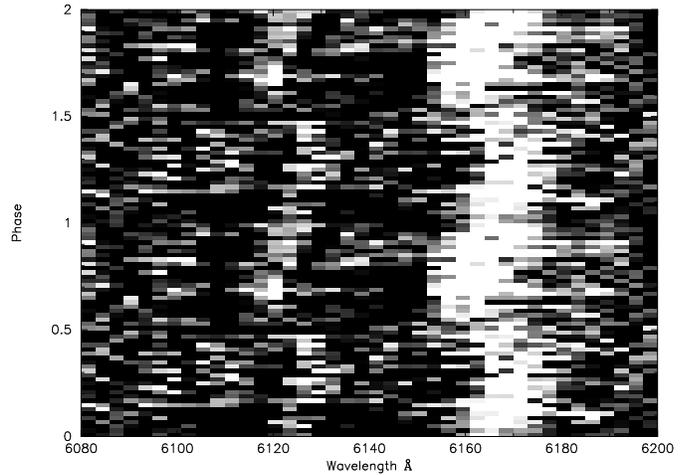}}
\caption[]{Trailed spectrogram of the Ca\,{\sc i} triplet between
6100-6200\AA. The Ca\,{\sc i} $\lambda$6162 is the strongest of the
three, and the others are located at $\lambda$6121 and
$\lambda$6102. This last one is only slightly visible. 
\label{fig:trailAbs}}
\end{figure}

In Fig.\ \ref{fig:trailAbs} we show a trailed spectrum of the region
around 6160\AA. We see that indeed there are absorption lines
present, at $\sim$6160\AA\, at $\sim$6120\AA\ and a trace can be seen
of a third line around 6100\AA.  These three wavelengths uniquely
identify this set of lines as the Ca\,{\sc i} triplet of
$\lambda$6102, $\lambda$6122 and $\lambda$6162.

\begin{figure}[htb]
\centerline{\psfig{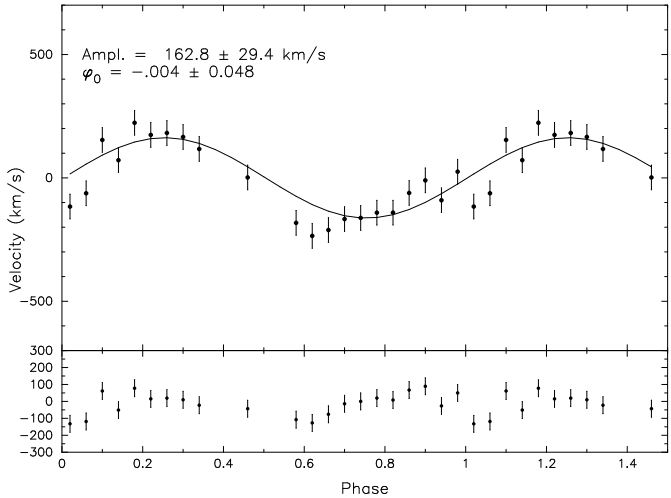}}
\caption[]{The radial velocity curve of the Ca\,{\sc i} absorption line
at 6162\AA. The phasing and amplitude of the radial velocity curve
correspond well with a place of origin on the secondary star. 
The decrease in the absorption strength as seen in Fig.\
\ref{fig:trailAbs} prevents a determination of the radial velocity
around phase 0.5. 
\label{fig:radvelsecrw}}
\end{figure}

We have determined the radial velocity curve of the secondary by
auto-correlating the absorption profile of the $\lambda$6162 line at
all phases with the profile between phase 0.65$<\varphi<$0.7 (Fig.\
\ref{fig:radvelsecrw}). A sinusoidal fit to the phase interval
0.1$<\varphi<$0.9 shows that the phasing of the line coincides with
that of the secondary star and the derived amplitude is consistent
with that derived by Smith et al. (1993) from the near infrared
Na\,{\sc i} lines.  The strength of the absorption lines reaches a
minimum between phase 0.45$<\varphi<$0.55, caused by irradiation. 

\section{Line light curves \label{sec:linelight}}

\begin{figure*}[htb]
\centerline{\psfig{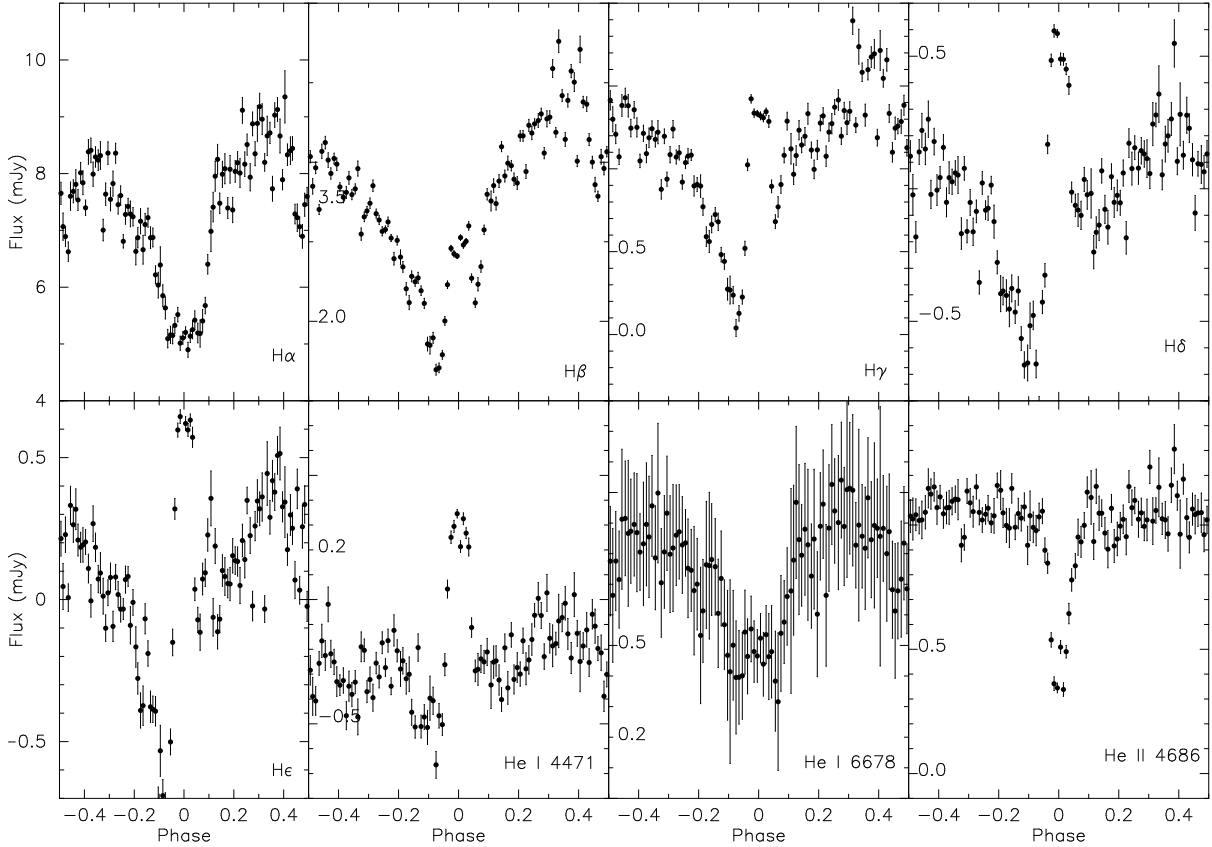}}
\caption[]{The light curves of the same lines as showed in Fig.\
\ref{fig:trailrwtri}. 
\label{fig:lightrwtri}}
\end{figure*}

In Fig.\ \ref{fig:lightrwtri} we show the light curves of the same lines
as shown in Fig.\ \ref{fig:trailrwtri} at a phase resolution of 0.01P$_{\rm
orb}$. All lines show a similar behaviour, with the possible
exception of \He2. The light curve of this line, however, is very
similar to that of the continuum. The decrease of light in the Balmer
lines during eclipse is very moderate and much less pronounced than that of the
continuum, which already indicates that most of the line emitting
region remains visible during eclipse. 

\begin{figure}[htb]
\centerline{\psfig{figure=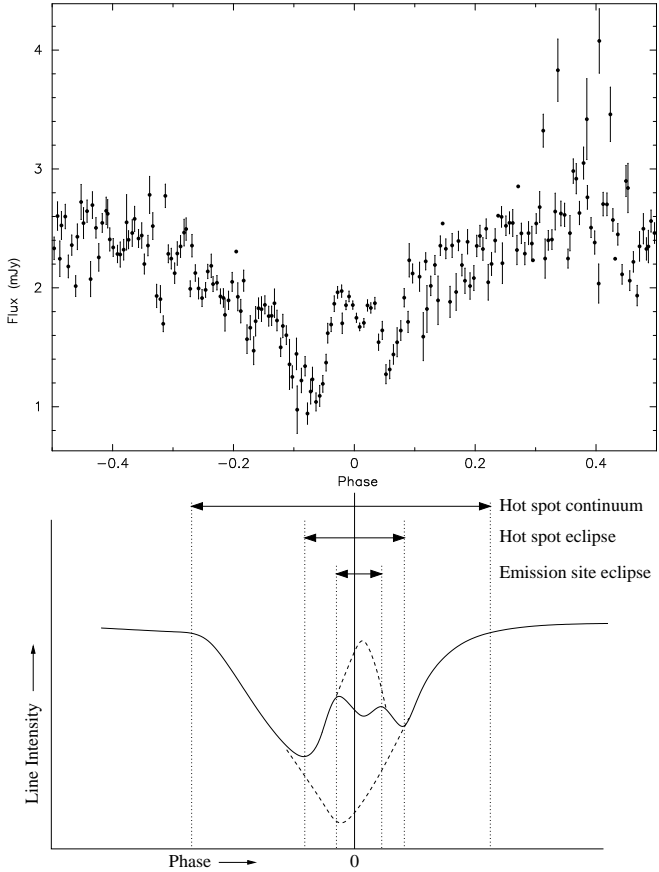,angle=0,width=8.8cm}}
\caption[]{The light curve of H$\beta$ at a phase resolution of
0.005P$_{\rm orb}$ (top), and a schematic view of the light curve,
indicating the different components (bottom).
\label{fig:lightHb}}
\end{figure}

To explain in more detail the behaviour of the lines, we show an
expanded view of the light curve of H$\beta$, at 0.005 phase
intervals, and a schematic view of the light curve in Fig.\ \ref{fig:lightHb}

Between phases 0.3$<\varphi<$0.6 the light curve is roughly
constant. At phase $\varphi\sim$0.65 a decrease in the line intensity
starts. This decrease is part of a very wide (0.6$<\varphi<$0.2),
slightly skewed, V-shaped feature, that is centered on
$\varphi\sim$0.95. This phase interval and its center coincides with
the visibility of a radiating hot-spot on the rim of a geometrically
thick accretion disk (e.g. as in IP Peg, Marsh, 1988; Groot,
1999). Marsh (1988) and Groot et al. (2001) have shown the hot-spot to
be typically around 10\,000 K, corresponding to a late B, early A-type
spectrum. This spectral type is characterized by very strong
absorption lines of the Balmer series. We therefore expect the
hot-spot region in RW Tri to be the source of strong Balmer (and He
{\sc i}) absorption lines. We conclude that the broad V-shaped feature
between 0.6$<\varphi<$0.2 is due to the visibility of these strong
absorption lines in the hot spot radiation.

Between phase 0.95$<\varphi<0.05$ the line intensity increases again,
and in the higher Balmer series (H$\delta$ and up) actually reaches
emission levels that are higher than outside eclipse. This
phase-interval coincides with the eclipse of the accretion disk, and
we conclude that the rise in line intensity is due to the fact that
the hot spot and accretion disk continuum, with its deep absorption
lines, is eclipsed during this phase interval. That the intensities in
the higher Balmer lines reach higher levels than outside eclipse is
because their line strength is a combination of isotropic emission
from an extended, optically thin, region, anisotropic absorption from
the hot spot {\sl and} isotropic absorption from the rest of the disk.
During eclipse both absorption components are occulted, leaving solely
the emission.

At phase interval $0.98<\varphi<0.02$ the line intensity decreases
slightly again. This is seen for H$\beta$ in Fig.\ \ref{fig:lightHb},
but the same behaviour is also shown by the other lines (as seen at
lower phase resolution in Fig.\ \ref{fig:lightrwtri}).  This
short-lasting decrease can only be due to the eclipse of the emission
site itself. The brief central dip is off-centered with the respect to
the line connecting the secondary star with the white dwarf, very
similar in phasing to the hot spot. We therefore conclude that the
main emission region of both the Balmer lines as well as the He{\sc
ii} 4686 line and the Bowen blend is centered on the hot spot region,
which was also concluded from the offset in the radial velocity
curves. 

\section{Spectral Eclipse Mapping \label{sec:emaprwtri}}

For the eclipse mapping procedure, we used the run-combined light
curve to obtain sufficient phase resolution and phase
coverage. Analysis of the light curves showed that the profile of the
eclipse of October 24 deviated in its shape from the rest of the
eclipse profiles, especially in the steepness of the egress
feature. This eclipse profile has therefore been excluded from the
run-combined average.

We have straightened the out-of-eclipse
profiles by fitting a spline curve to the phases $\varphi<$--0.12 and
$\varphi>$0.12, and also brought the light curves to a common scale,
which was chosen to be the brightness of RW Tri on the first night at
AB=13.0 at 4500\AA.

\begin{figure}[htb]
\centerline{\psfig{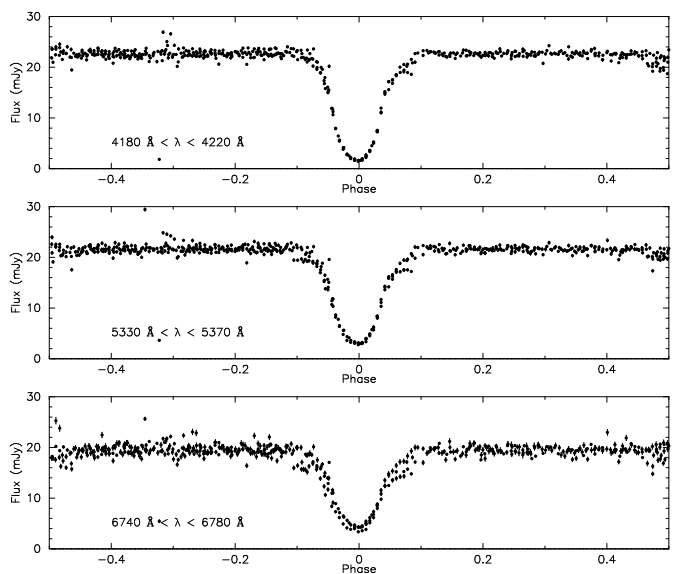}}
\caption[]{The run-combined straightened light curves of RW Tri in three
wavelength regions centered on 4200\AA, 5350\AA\ and 6760\AA. 
\label{fig:lightnarrow}}
\end{figure}

The phase-resolved spectra of RW Tri have been divided in 80 narrow band light
curves, each 40\AA\ wide, except around the spectral lines, which were
taken as single bins. In Fig.\ \ref{fig:lightnarrow} we show the
corrected light curves in three narrow band wavelength regions,
spanning the wavelength range covered. We see that, despite
the variation in the level of the out-of-eclipse light, the eclipse
profiles themselves are very similar. The amount of scatter on the
narrow-band light curves increases when going to the red. This is most
likely caused by an inaccurate slit correction in the red part of the
disk. 

For the eclipse mapping reconstruction we have used a 51$\times$51 pixel map,
phasebins of 0.005 in phase and the system parameters as given in
Table\ \ref{tab:sysrwtri}.

\subsection{Disk size}

To measure the size of the accretion disk at different wavelengths we
have used the distance (R$_{0.1}$) 
where the intensity on the disk has fallen to
10\% of the central intensity. This measure was used by Rutten et
al. (1992) to compare the relative sizes of the disks in six different
novalike systems. We find from our eclipse maps that R$_{0.1}$ =
0.25$\pm$0.10\RL1 at 4420\AA. The rather large error is caused by a
relatively flat run of the reconstructed intensity with radial
distance at this wavelength. For 6270\AA\ the disk size has increased
to R$_{0.1}$ = 0.45$\pm$0.05\RL1. Both values are comparable to the values
found by Rutten et al. (1992) at 4410\AA\ (R$_{0.1}$=0.28$\pm$0.03\RL1)
and 8010\AA\ (R$_{0.1}$=0.43$\pm$0.03\RL1). 

\subsection{Accretion disk annuli spectra \label{sec:diskspectra}}

\begin{figure}[htb]
\centerline{\psfig{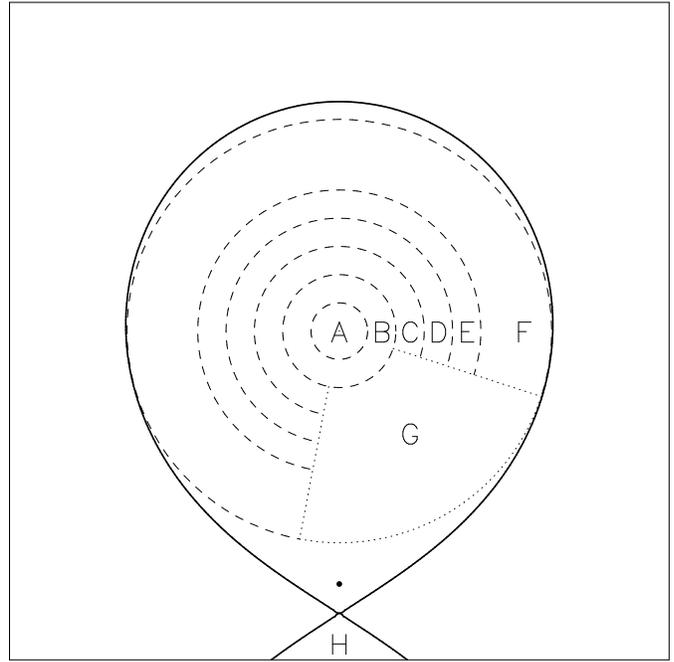}}
\caption[]{Schematic view of the white dwarf Roche lobe, showing the
subdivision of the Roche lobe in seven regions, labeled `A'-`G'. The
uneclipsed light is denoted by region `H', tentatively placed on
the secondary.
\label{fig:rings_rwtri}}
\end{figure}

\begin{figure*}[htb]
\centerline{\psfig{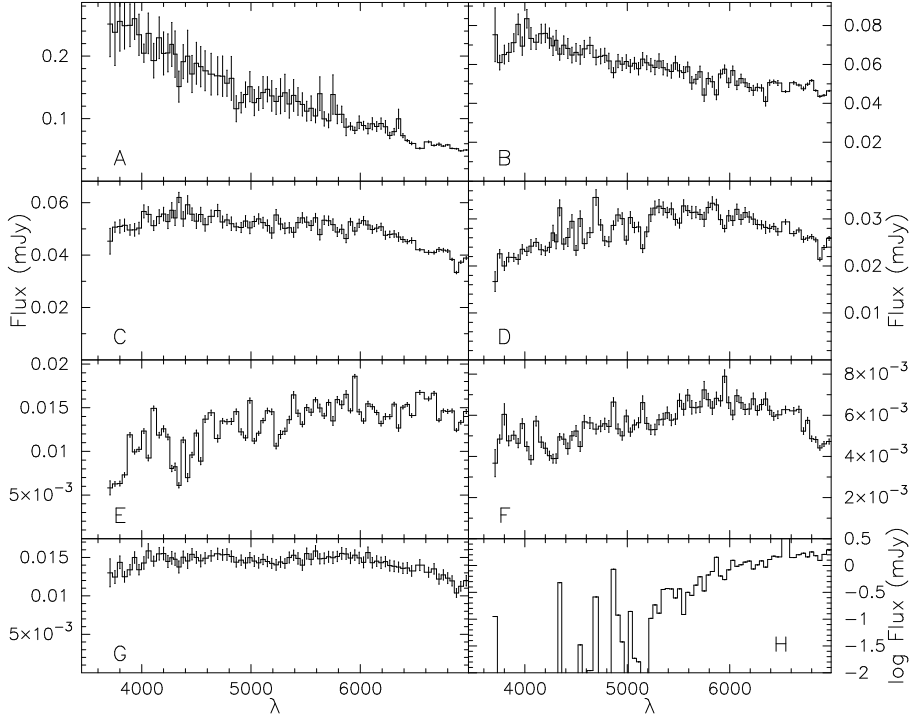}}
\caption[]{The reconstructed spectrum of the accretion disk in RW Tri
in the regions defined in Fig.\ \ref{fig:rings_rwtri}. Fluxes are
given per surface tile element (each 1.6$\times$10$^{18}$ cm$^2$
including the foreshortening at $i$=75\degr). 
We see that the slope of the spectrum
changes dramatically from very blue in the inner parts, to red in the
outer parts. The hot-spot area (region `G') is more blue than the rest
of the outer disk. The fluxes of the uneclipsed light component
(region `H') are plotted on a logarithmic scale, all others on a
linear scale. \label{fig:radialspectrum_rw}}
\end{figure*}

We have defined seven regions in the accretion disk of RW Tri, labeled
`A' through `G' at radii of 0.1, 0.2, 0.3, 0.4, 0.5, 0.7 \RL1 and a
hot-spot region ranging from 0.2-0.7 \RL1 and 0.8$<\varphi<$0.03
 (Fig.\ \ref{fig:rings_rwtri}). The spectra
of these regions are shown in Fig.\ \ref{fig:radialspectrum_rw}. We
see that there is a strong change in the slope of the spectrum 
from blue near the white dwarf to red in the outer parts of the accretion
disk. No clear signature of any spectral lines (apart perhaps from the
Balmer series in absorption in region `A') is seen.  Not surprisingly
the non-eclipsed light (region `H') shows the spectral lines (\Ha,
He{\sc i} $\lambda$5875, H$\beta$, \He2, H$\gamma$ and H$\delta$) in
emission on top of a continuum level that rises to the red. 

\subsection{Distance to RW Tri \label{sec:distance}}

In order to convert the reconstructed fluxes to specific intensities,
from which temperatures can be derived by e.g. blackbody fitting, it
is imperative to have a good estimate of the distance to the system. The
distance of RW Tri was recently determined by parallax measurements
using the HST Fine Guidance Sensor to be 341$\pm$35 pc (McArthur et
al., 1999). 

The distance of RW Tri can also be estimated by allowing the
distance as well as the temperature to vary in blackbody fits to
the reconstructed accretion disk spectra. Blackbody fitting in the
wavelength region 4000-6200\AA\ gives a
distance to RW Tri of 330$\pm$40 pc, in excellent agreement with the
parallax measurements and also with the estimate of the fractional
contribution of the secondary by Rutten et al. (1992; 330 pc). We refer to
McArthur et al. (1999) for a comparison with other distance estimates. 
We will use the value of 330 pc in the further analysis. 

\subsection{The radial temperature profile \label{sec:radtemp}}

To determine the radial temperature profile of the accretion disk we
have used the wavelength region of 4000\AA-6200\AA, from which the
emission lines have been eliminated. The blue cut-off has been chosen to
avoid any influence of the Balmer jump and the red cut-off has been
chosen because trial blackbody fits showed that the reconstructed
intensities at these wavelengths strongly deviated from the expected
values based on the trend in the bluer part of the wavelength which
generally agreed well with the blackbody fits. 

\begin{figure}[htb]
\centerline{\psfig{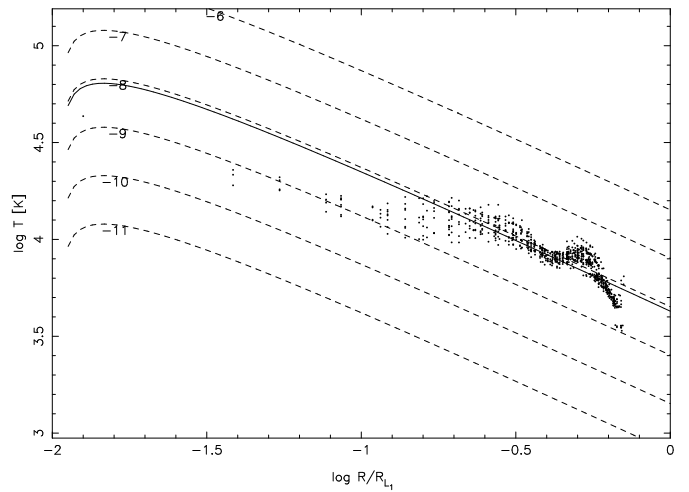}}
\caption[]{The radial temperature profile of RW Tri, as deduced from
the spectral eclipse mapping. The dashed lines show
the theoretical prediction of the radial temperature profile based on
the theory of optically thick, steady state accretion disks.
\label{fig:mdot_rwtri}}
\end{figure}

After the determination of the distance as described in the previous
paragraph (330$\pm$40 pc), we fixed the distance at this value and
fitted only for the temperature. For this we have taken the spectrum
at each pixel (in specific intensities) and fitted a temperature to
this spectrum.  We show the radial temperature profile of the
accretion disk in RW Tri in Fig.\ \ref{fig:mdot_rwtri}, which also
shows the theoretical predictions for the radial temperature profile
based on optically thick, steady state accretion disks. To calculate
these theoretical profiles we have used the white dwarf mass as given
in Tab.\ \ref{tab:sysrwtri} and the white dwarf mass-radius relation
of Nauenberg (1972).

To determine the position of the hot-spot we have taken the ratio of
the continuum intensity maps at 4060\AA\ and one at 6270\AA\ to
maximize the constrast between the and blue hot-spot and the cooler
outer disk. This ratio shows the position of the hot-spot to be
centered on ($r,\varphi$) of (0.5\RL1,0.875).

\section{Discussion \label{sec:discussion}}

\subsection{Continuum radiation from the eclipse maps: mass-transfer rate}

\begin{figure}[htb]
\centerline{\psfig{figure=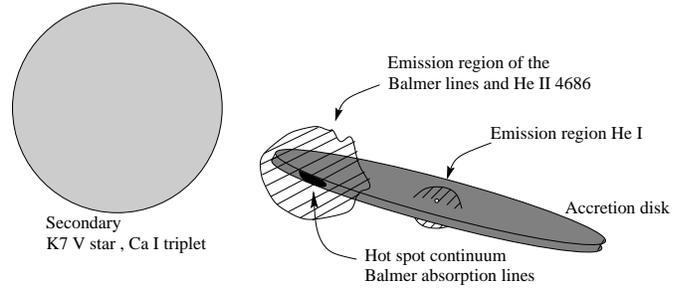,width=8.8cm,angle=-90}}
\caption[]{Schematic view of the structure of RW Tri and the location
of the areas producing the spectral lines discussed in the
text. \label{fig:balmer}}
\end{figure}

From the wealth of information RW Tri displays in our dataset we
construct a conceptual model of the structure of RW Tri 
(see Fig.\ \ref{fig:balmer}). 

The accretion disk, although showing substantial variation on its
radial temperature profile, on average follows the standard, steady
state, T $\propto$R$^{-3/4}$, profile predicted from theory and
observed in RW Tri before (Rutten et al., 1992), up until a radius of
0.15 \RL1 from the white dwarf, where the temperature profile flattens
off. 
Comparing our temperature profile with those
derived by Rutten et al. (1992) on the basis of four-colour photometry
and by Horne \& Stiening (1985) on the basis of brightness temperature
estimates of their B band photometry, we see that our profile already
levels off at a larger distance from the white dwarf (at
$\sim$0.15\RL1 here and at $\sim$0.06\RL1 in Rutten et al., 1992 and
Horne \& Stiening, 1985.)

From the radial temperature profile between 0.2 - 0.56 R$_{\rm L_1}$
we deduce a mass-transfer rate through the disk of $\sim$10$^{-8}$
M$_{\odot}$ yr$^{-1}$.  This mass transfer rate is somewhat higher
than in Rutten et al. (1992), but this is partly caused by the lower
distance (270 pc) used in that study.

\subsection{Line profiles: velocities and emission sites}

The strong presence of emission lines in the uneclipsed light
indicates that they are formed out of the plane of the disk. 

The fact that during eclipse an emission component 
appears in He {\sc i} $\lambda$4026 and $\lambda$4471, suggests that
the absorption we see during most of the orbit is a combination of
absorption and emission, where the absorption is caused by material
close to the disk, which is back-lit by the white dwarf and
surroundings, and the emission is caused by material higher above the
disk, which is, in our line-of-sight, not back-lit by material behind
it. This would also explain why the He {\sc i} $\lambda$6678 line is in
emission, since this is the line with the lowest excitation level and
therefore presumably the largest emission region.

In the trailed
spectra the emission lines are characterized by a high and low velocity
component.  The phasing of the high-velocity component indicates that
the component originates on the line of centers and on the white dwarf
side of the center-of-mass.  The amplitude is too low to be connected
with any rotational velocity in the accretion disk (which is assumed
to have a Keplerian rotation field, with a minimum velocity of
$\sim$500 \kms\ in the outer regions of the disk). Considering both
these two facts (no phase lag and an amplitude that is too low for
rotation of the accretion disk) it appears that the white dwarf itself
or its immediate surroundings is the place of origin of the high
velocity component. In this case the (uncertain) amplitude of 330 \kms\
would reflect the orbital velocity of the white dwarf.

The radial velocity curve of the low-velocity component shows a
significant offset with respect to the line of centers, and this phase
lag puts is at the position where the hot-spot is located. The
amplitude also coincides with the velocity of the hot-spot region in
the binary frame. This means that both the gas at the emission site
of the low-velocity component, as well as that at the emission site of
the high-velocity component must be decoupled from the Keplerian flow
in the disk. 

Overall it appears that the spectral lines are dominated by two
emission regions that both have appreciable vertical extension with
respect to the disk. The most dominant of these region is located at
the hot spot (located at (r,$\varphi$) = 0.5\RL1, 0.875), as evidenced
by the phase lag, radial velocity amplitude and light curve of the
Balmer and He {\sc i} emission lines. Also the \He2 line (and the
C{\sc iii}/N{\sc iii} Bowen blend) is formed in this region, as
evidenced by the identical radial velocity curves of \Ha and \He2. The
marked difference in the \Ha and \He2 line light curves, where the \Ha
light curve shows a very broad absorption lasting from
0.6$<\varphi<$0.2, whereas the \He2 light curve is very similar to
that of the continuum is caused by the fact that the hot spot is too
cool to cause appreciable anisotropic \He2 absorption, and the \He2
light curve therefore shows `just' the eclipse by the secondary
star. Also, the \He2 emission region is much smaller than the Balmer
region, since no central emission is seen at mid-eclipse.

\subsection{SW Sex behaviour}

During our observations RW Tri showed some features that are commonly
used as identifiers of the SW Sex sub-class of novalike CVs (see
Thorstensen et al., 1991 and Groot et al., 2001): single-peaked
emission lines, low amplitude radial velocity curves of the main
Balmer lines, phase lags of the Balmer lines, shallow eclipses of the
Balmer lines, a flat radial temperature profile and \He2
emission. Although some of these indicators may be more hinted at in
our observations than firmly proven, the combined occurence of these
indicators raises the question whether RW Tri should be considered to
be an SW Sex star. 

Groot et al. (2001) showed that in recent observations of SW Sex in a
low state, this system lacked some of the SW Sex features, e.g. the
phase 0.5 absorption. Combined with the RW Tri observations presented
here it appears that the boundary between UX UMa-, RW Tri- and SW
Sex-like novalikes (see Warner 1995 for definitions) is vague and
depends on the brightness of the system at the moment of
observation. It appears that the behaviour that is `standard' for the
sub-classes are the extremes of a sliding scale, most likely depending
on the mass-transfer rate, where the SW Sex behaviour becomes more
prominent with increasing mass-transfer rate.

\begin{acknowledgements}
PJG whishes to thank Alex de Koter for pleasant discussions on the
line formation mechanisms in non-LTE regimes, and Claudio Moreno and
the staff of the ING observatory for their hospitality during various
visits. We would also like to thank the referee, dr. Bruch, for his
very useful comments and Janet Drew for reading and commenting on the
manuscript.  PJG acknowledges partial support by NWO Spinoza grant
08-0 to E.P.J. van den Heuvel and an NWO-VIDI grant 639.042.201 to
P.J. Groot. The Isaac Newton Telescope is operated by the Isaac Newton
Group in the Spanish Observatorio del Roque de los Muchachos of the
Instituto de Astrof\'{\i}sica de Canarias.
\end{acknowledgements}

\end{document}